\documentclass[12pt,aps,onecolumn,nobibnotes]{revtex4}

\usepackage{amsmath}
\usepackage{amssymb}

\def\l{\lambda}
\def\bsq{~~$\blacksquare$}
\def\bbz{Z\!\!\!Z}
\def\({\left(}\def\){\right)}
\def\hH{{\hat H}} 
\def\x{\xi}\def\y{\eta}
\def\rg{\rangle} 
\def\witl{\widetilde{\vert \L\rg}}
\def\bbn{I\!\!N}
\def\r{\rho}\def\om{\omega}
\def\half{{\textstyle{\frac{1}{2}}}}
\def\trh{{\textstyle{\frac{3}{2}}}}
\def\td{\tilde d}
\def\riga{-\kern-5pt - \kern-5pt -}
\def\white#1{\mathop{\bigcirc}\limits_{#1}}
\def\white#1{\mathop{\circ}\limits_{#1}}
\def\black#1{\mathop{\bullet}\limits_{#1}}
\def\downcirc#1{\mathop{\circ}\limits_{#1}}

\def\o{{\bar 0}} \def\I{{\bar 1}}
\def\a{\alpha}
\def\b{\beta}
\def\d{\delta}

\def\D{\Delta} \def\L{\Lambda}
\def\bac{{C\kern-5.5pt I}}
\def\bbc{{C\kern-8pt I}}

\def\be{\begin{equation}}
\def\eqn{\begin{equation}\label}
\newcommand{\ee}{\end{equation}}

\def\bea{\begin{eqnarray}}
\def\eqnn{\bea\label}
\def\eea{\end{eqnarray}}
\def\nn{\nonumber}

\newcommand{\eqna}[1]{\begin{subequations} \label{#1}
\begin{eqnarray}}
\def\eena{\end{eqnarray}
\end{subequations}}

\def\vr{\vert} \def\bbr{I\!\!R}

\def\nt{\noindent}
\def\nl{\hfil\break}
\def\np{\vfil\eject}

  \def\cc{{\cal C}}
  
\def\cg{{\cal G}} \def\ch{{\cal H}} 
 \def\ck{{\cal K}} 
  
\def\cp{{\cal P}}

\begin{document}

\title{Positive Energy Unitary
Irreducible Representations of the Superalgebras $osp(1|2n,\bbr)$}

\author{V.K. Dobrev}
\thanks{Permanent address:
Institute of Nuclear Research and Nuclear Energy,
Bulgarian Academy of Sciences, Sofia, Bulgaria.}
\affiliation
{School of Informatics,
University of Northumbria,\\ Newcastle-upon-Tyne NE1 8ST, UK}

 \author{R.B. Zhang}
\affiliation{School of Mathematics and Statistics,\\
University of Sydney,\\ Sydney, New South Wales 2006, Australia}

\begin{abstract}
We give the classification of the positive energy
(lowest weight) unitary irreducible representations of
the superalgebras ~$osp(1|2n,\bbr)$.
\end{abstract}

\maketitle


\section{Introduction}

Recently, superconformal field theories in various dimensions are
attracting more interest, in particular, due to their duality to
AdS supergravities, cf. [1-58] and references therein.
Until recently only those for ~$D\leq 6$~ were studied since in
these cases the relevant superconformal algebras satisfy \cite{Nahm}
the Haag-Lopuszanski-Sohnius theorem \cite{HLS}.
Thus, such classification was known only for the ~$D=4$~
superconformal algebras ~$su(2,2/N)$ \cite{FF} (for $N=1$),
\cite{DPm,DPu,DPf,DPp} (for arbitrary $N$). More recently, the
classification for ~$D=3$ (for even $N$), $D=5$, and $D=6$ (for
$N=1,2$)~ was given in \cite{Min} (some results are conjectural),
and then the $D=6$ case (for arbitrary $N$) was finalized in \cite{Dosix}.

On the other hand the applications in string theory require the
knowledge of the UIRs of the conformal superalgebras for ~$D>6$.
Most prominent role play the superalgebras $osp(1\vr\,2n)$,
cf. their applications in, e.g.,
\cite{Tow,HW,PHNW,Gun,Luk,BvP,DFLV,Ban,SWK,BCB}.
Initially, the superalgebra $osp(1\vr\,32)$
was put forward for $D=10$ \cite{Tow}. Later it was realized that
$osp(1\vr\,2n)$ would fit any dimension, though they are minimal
only for $D=3,9,10,11$ (for $n=2,16,16,32$, resp.) \cite{DFLV}.
In all cases we need to find first the UIRs of
~$osp(1\vr\, 2n,\bbr)$. This can be done for general $n$.
Thus, in this paper we treat the UIRs of
~$osp(1\vr\, 2n,\bbr)$~ only while the implications for
conformal supersymmetry for $D=9,10,11$ shall be treated
in a follow-up paper.

\section{Representations of the superalgebras $\ osp(1\vr\, 2n)\ $
and $\ osp(1\vr\, 2n,\bbr)$}

\subsection{The setting}

\nt
Our basic references for Lie superalgebras are \cite{Kab,Kc}.
The conformal superalgebras in $D=9,10,11$ are ~$\cg ~=~
osp(1\vr\, 2n,\bbr)$, $n=16,16,32$, resp., cf. \cite{Tow,DFLV}.
The even subalgebra of ~$osp(1\vr\, 2n,\bbr)$~ is the algebra
~$sp(2n,\bbr)$ with maximal compact subalgebra
~$\ck = u(n) \cong su(n) \oplus u(1)$.
The algebra ~$sp(2n,\bbr)$~ contains the conformal algebra ~$\cc
~=~ so(D,2)$, while ~$\ck$~ contains the maximal compact subalgebra
~$so(D)\oplus so(2)$~ of ~$\cc$,
$so(2)$ being idenitified with the $u(1)$ factor of $\ck$.

We label the relevant representations of ~$\cg$~
by the signature:
\eqn{sgn}\chi ~=~ [\, d\,;\,a_1\,,...,a_{n-1}\,] \end{equation}
where ~$d$~ is the conformal weight, and ~$a_1,...,a_{n-1}$~
are non-negative integers which are Dynkin labels of the
finite-dimensional UIRs of the subalgebra $su(n)$ (the simple
part of $\ck$).

Our aim is to classify the UIRs of ~$\cg$~
following the methods used for the $D=4,6$ conformal superalgebras,
cf. \cite{DPm,DPu,DPf,DPp},\cite{Dosix}, resp. The main tool is an adaptation
of the Shapovalov form on the Verma modules ~$V^\chi$~ over the
complexification ~$\cg^\bac ~=~ osp(1\vr\, 2n)$~ of ~$\cg$.

\subsection{Verma modules}

\nt
To introduce Verma modules we use the standard
triangular decomposition:
\eqn{trig} \cg^\bac ~=~ \cg^+ \oplus \ch \oplus \cg^- \end{equation}
where $\cg^+$, $\cg^-$, resp., are the subalgebras corresponding
to the positive, negative, roots, resp., and $\ch$ denotes the
Cartan subalgebra.

We consider lowest weight Verma modules,
so that ~$V^\L ~ \cong U(\cg^+) \otimes v_0\,$,
 where ~$U(\cg^+)$~ is the universal enveloping algebra of $\cg^+$,
and ~$v_0$~ is a lowest weight vector $v_0$ such that:
\eqnn{low}
 Z \ v_0\ &=&\ 0 \ , \quad Z\in \cg^- \nn\\
 H \ v_0 \ &=&\ \L(H)\ v_0 \ , \quad H\in \ch \ .\eea
Further, for simplicity we omit the sign
~$\otimes \,$, i.e., we write $p\,v_0\in V^\L$ with $p\in U(\cg^+)$.

The lowest weight $\L$ is characterized by its values on the
simple roots of the superalgebra. In the next subsection we
describe the root system.

\subsection{Root systems}

\nt
We recall some facts about ~$\cg^\bac ~=~ osp(1\vr\, 2n)$
(denoted $B(0,n)$ in \cite{Kab}).
Their root systems are given in terms of
~$\d_1\,\dots,\d_{n}\,$, ~$(\d_i,\d_j) ~=~
\d_{ij}\,$, ~$i,j=1,...,n$.
The even and odd roots systems are \cite{Kab}:
\eqn{dmn}
\D_\o ~~=~~ \{ \pm\d_i\pm\d_j\ , ~1\leq i< j\leq n\ ,
~~\pm 2\d_i\ , ~1\leq i \leq n \} ~,
\qquad \D_\I ~~=~~ \{ \pm\d_i\ , ~1\leq i \leq n
\} \end{equation}
(we remind that the signs ~$\pm$~ are not correlated).
We shall use the following distinguished
simple root system \cite{Kab}:
\eqn{ssdm}
 \Pi ~=~ \{\, \d_1-\d_2 \, ,\,
 , \dots, \d_{n-1}- \d_{n} \, ,\, \d_n\, \} \ , \end{equation}
or introducing standard notation for the simple roots:
\eqnn{ssdma}
\Pi ~&=&~ \{\,\a_1\,,...,\,\a_{n} \, \}\ , \\
&&\a_{j} ~=~ \d_{j}-\d_{j+1}\ , \quad
j=1,...,n-1 \ , \quad \a_{n} ~=~ \d_{n}\ . \nn\eea
The root ~$\a_n = \d_n$~ is odd, the other simple roots
are even.
The Dynkin diagram is:
\eqn{dynk}
\vbox{\offinterlineskip\baselineskip=10pt
\halign{\strut#
\hfil
& #\hfil
\cr &\cr
&\ $\white{{1}}
\riga \cdots \riga
\white{{n-1}} =\kern-2pt\Longrightarrow
\black{n}$
\cr }} \end{equation}
The black dot is used to signify that the simple odd root is not
nilpotent, otherwise a gray dot would be used \cite{Kab}. In fact, the
superalgebras ~$B(0,n) = osp(1\vr\, 2n)$~ have no nilpotent generators
unlike all other types of basic classical Lie superalgebras \cite{Kab}.

The corresponding to ~$\Pi$~ positive root system is:
\eqn{psdm}\D_\o^+ ~=~ \{ \d_i\pm\d_j\ , ~1\leq i< j\leq n\ ,
 ~~2\d_i \ , ~1\leq i \leq n \} ~,
\qquad \D_\I^+ ~=~ \{ \d_i \ , ~1\leq i \leq n \} \end{equation}
We record how the elementary functionals are expressed through
the simple roots:
\eqn{back} \d_k ~=~ \a_k + \cdots + \a_n \ .\end{equation}

The even root system ~$\D_\o$~ is the root system of the rank $n$
complex simple Lie algebra ~$sp(2n)$, with ~$\D^+_\o$~ being its
positive roots. The simple roots are:
\eqnn{ssdmb}
 \Pi_0 ~&=&~ \{\, \d_1-\d_2 \, ,\,
 , \dots, \d_{n-1}- \d_{n} \, ,\, 2\d_n\, \} ~=~
\{\,\a^0_1\,,...,\,\a^0_{n} \, \}\ , \\
&&\a^0_{j} ~=~ \d_{j}-\d_{j+1}\ , \quad
j=1,...,n-1 \ , \qquad \a^0_{n} ~=~ 2\d_{n}\ .
\nn\eea
The Dynkin diagram is:
\eqn{dynko}
\vbox{\offinterlineskip\baselineskip=10pt
\halign{\strut#
\hfil
& #\hfil
\cr &\cr
&\ $\downcirc{{1}} \riga \dots
\riga\downcirc{{n-1}} \Longleftarrow\kern-2pt=
\downcirc{{n}}$
\cr }}
\end{equation}

The superalgebra ~$\cg ~=~ osp(1\vr\, 2n,\bbr)$~ is a split real form
of $osp(1\vr\, 2n)$ and has the same root system.

\subsection{Lowest weight through the signature}

\nt
Since we use a Dynkin labelling, we have the
following relation with the signature ~$\chi$~ from (\ref{sgn}):
\eqnn{rrr}
(\L , \a_k^\vee ) ~=&&~
-\, a_k \ , ~~~ k<n\ , \nn\\
&& ~~~\td \ , ~~~k=n\ , \eea
where ~$\a_k^\vee \equiv 2\a_k/(\a_k,\a_k)$, and
~$\td$~ differs from the conformal weight ~$d$~ as
explained below. The minus signs in the first row
are related to the fact that we work with
lowest weight Verma modules (instead of the highest weight modules
used in \cite{Kc}{}) and to Verma module reducibility
w.r.t. the roots ~$\a_k$ (this is explained in detail in \cite{DPf}).
The value of ~$\td$~ is a matter of
normalization so as to correspond to some known cases.
Thus, our choice is:
\eqn{cftw} \td ~=~ 2 d + a_1 + \cdots + a_{n-1} \ . \end{equation}

Having in hand the values of ~$\L$~ on the basis we can
recover them for any element of ~$\ch^*$.
In particular, for the values on the elementary functionals
we have using (\ref{back}),(\ref{rrr}) and (\ref{cftw}):
\eqn{vell} (\L ,\d_j ) ~=~ d\
+\ \half (a_1 + \cdots + a_{j-1} - a_j - \cdots - a_{n-1} ) \ .
\end{equation}
Using (\ref{rrr}) and (\ref{cftw}) one can write easily ~$\L =
\L(\chi)$~ as a linear combination of the simple roots or of the
elementary functionals ~$\d_j\,$, but this is
not necessary in what follows.
We shall need only ~$(\L , \b^\vee)$~ for all positive roots $\b$
and from (\ref{vell}) we have:
\eqnn{lorr}
(\L , (\d_i-\d_j)^\vee ) ~&=&~ (\L , \d_i-\d_j ) ~=~
- a_i - \cdots - a_{j-1} \nn\\
(\L , (\d_i+\d_j)^\vee ) ~&=&~ (\L , \d_i+\d_j ) ~=~
2d\ +\ a_1 + \cdots + a_{i-1} - a_j - \cdots - a_{n-1} \qquad
\nn\\
(\L , \d_i^\vee ) ~&=&~ (\L , 2\d_i ) ~=~
2d \ +\ a_1 + \cdots + a_{i-1} - a_i - \cdots - a_{n-1}
\nn\\
(\L , (2\d_i)^\vee ) ~&=&~ (\L , \d_i ) ~=~
d \ +\ \half ( a_1 + \cdots + a_{i-1} - a_i - \cdots - a_{n-1} )
\eea

\subsection{Reducibility of Verma modules}

\nt
Having established the relation between ~$\chi$~ and ~$\L$~
we turn our attention to the question of reducibility.
A Verma module ~$V^\L$~ is reducible w.r.t.
the positive root ~$\b$~ iff the following holds \cite{Kc}:
\eqn{odr} (\r - \L, \b^\vee) = m_\b\ , \qquad \b \in \Delta^+
\ , \quad m_\b\in\bbn \ , \end{equation}
where ~$\r\in\ch^*$~ is the very important in representation
theory element given by the
difference of the half-sums $\r_\o\,,\r_\I$
of the even, odd, resp., positive roots (cf. (\ref{psdm}):
\eqnn{hlf} \r ~&\doteq&~ \r_\o - \r_\I
~=~ (n-\half)\d_1 + (n-\trh) \d_2 + \cdots + \trh \d_{n-1} +
\half \d_n \ , \\
&&\r_\o\ =\ n \d_1 + (n-1) \d_2 + \cdots +
2\d_{n-1} + \d_n \ , \nn\\
&&\r_\I \ =\ \half (\d_1 + \cdots + \d_n)\ . \nn\eea

To make (\ref{odr}) explicit we need first the values of
 ~$\r$~ on the positive odd roots:
\eqn{ror} (\,\r\, , \d_i\, ) ~=~ n-i +\half \ . \end{equation}
Then for ~$(\r , \b^\vee)$~ we have:
\eqnn{rorr}
(\r , (\d_i-\d_j)^\vee ) ~&=&~ j-i \ ,
\nn\\
(\r , (\d_i+\d_j)^\vee ) ~&=&~ 2n -i-j +1 \ ,
\nn\\
(\r , \d_i^\vee ) ~&=&~ 2n-2i+1 \ ,
\nn\\
(\r , (2\d_i)^\vee ) ~&=&~ n-i+\half \ . \eea
Naturally, the value of $\r$ on the simple roots is 1:
~$(\r,\a^\vee_i)=1$, $i=1,...,n$.

Consecutively we find that the Verma module ~$V^{\L(\chi)}$~
is reducible if one of the following relations holds (following
the order of (\ref{lorr}) and (\ref{rorr}):
\eqna{redd}
\bbn ~\ni ~m^-_{ij} ~&=&~ j-i + a_i + \cdots + a_{j-1} \ ,\\
\bbn ~\ni ~m^+_{ij} ~&=&~ 2n -i-j +1 + a_j + \cdots + a_{n-1}
- a_1 - \cdots - a_{i-1} -2d\ , \\
\bbn ~\ni ~m_i ~&=&~ 2n-2i+1 + a_i + \cdots + a_{n-1} - a_1 +
\cdots - a_{i-1} - 2d\ , \\
\bbn ~\ni ~m_{ii} ~&=&~ n-i+\half(1 + a_i + \cdots + a_{n-1} - a_1 +
\cdots - a_{i-1}) - d \ . \eena
Note that ~$m_i ~=~ 2m_{ii}\,$, thus, whenever (\ref{redd}{d}) is
fulfilled also (\ref{redd}{c}) is fulfilled.

If a condition from (\ref{redd}) is fulfilled then ~$V^\L$~ contains a
submodule which is a Verma module ~$V^{\L'}$~ with shifted weight
given by the pair ~$m,\b$~: ~$\L' ~=~ \L + m\b$. The embedding of
~$V^{\L'}$~ in ~$V^\L$~ is provided by mapping the lowest weight vector
~$v'_0$~ of ~$V^{\L'}$~ to the singular vector ~$v_s^{m,\b}$~
in ~$V^\L$~ which is completely determined by the conditions:
\eqnn{lowp}
 X \ v_s^{m,\b}\ &=&\ 0 \ , \quad X\in \cg^- \ , \nn\\
 H \ v_s^{m,\b} \ &=&\ \L'(H)\ v_0 \ , \quad H\in \ch \ ,
~~~\L' ~=~ \L + m\b\ .\eea
Explicitly, ~$v_s^{m,\b}$~ is given by a polynomial in the positive
root generators:
\eqn{sing} v_s^{m,\b} ~=~ P^{m,\b} \,v_0 \ , \quad P^{m,\b}\in
U(\cg^+)\ . \end{equation}
Thus, the submodule of ~$V^\L$~ which is isomorphic to ~$V^{\L'}$~ is given
by ~$U(\cg^+)\, P^{m,\b} \,v_0\,$.

Here we should note that we may eliminate the reducibilities and
embeddings related to the roots ~$2\d_i\,$. Indeed, let (\ref{redd}{d})
hold, then the corresponding singular vector
~$v_s^{m_{ii},2\d_i}$~ has the properties prescribed by
(\ref{lowp}) with ~$\L' ~=~ \L + m_{ii}\,2\d_i\,$. But as we mentioned
above in this situation also (\ref{redd}{c})
holds and the corresponding singular vector
~$v_s^{m_i,\d_i}$~ has the properties prescribed by
(\ref{lowp}) with ~$\L'' ~=~ \L + m_i\,\d_i\,$. But due to
the fact that ~$m_i ~=~ 2m_{ii}$~ it is clear that ~$\L''=\L'$,
which means that the singular vectors ~$v_s^{m_{ii},2\d_i}$~
and ~$v_s^{m_i,\d_i}$~ coincide (up to nonzero multiplicative
constant). On the other hand if (\ref{redd}{c}) holds with $m_i$ being
an odd number, then (\ref{redd}{d}) does not hold (since ~$m_{ii} =
m_i/2$ is not integer).

Further, we notice that all reducibility conditions in (\ref{redd}{a}) are
fulfilled. In particular, for the simple roots from those
condition (\ref{redd}{a}) is fulfilled with
~$\b\to\a_i=\d_i-\d_{i+1}\,$, $i=1,...,n-1$
and ~$m^-_i ~\equiv~ m^-_{i,i+1} ~=~ 1 + a_i\,$. The
corresponding submodules ~$I^\L_i ~=~ U(\cg^+)\,v^i_s\,$,
where ~$\L_i ~=~ \L + m^-_i \a_i$~ and
~$v^i_s ~=~ (X^+_i)^{1+a_i}\, v_0\,$, where ~$X^+_i$~ are
the root vectors of the these simple roots. These submodules generate an
invariant submodule which we denote by ~$I^\L_c\,$. Since these
submodules are nontrivial for all our signatures
instead of ~$V^\L$~ we shall consider the factor-modules:
\eqn{fcc} F^\L ~=~ V^\L\, /\, I^\L_c \ . \end{equation}
We shall denote the lowest weight vector of $F^\L$ by
~$\widetilde{\vr \L\rg}$~ and the singular vectors above become null
conditions in $F^\L$~:
\eqn{nullm}
(X^+_i)^{1+a_i}\, \widetilde{\vr \L\rg} ~=~ 0 \ . \quad i=1,...,n-1.
\end{equation}

If the Verma module ~$V^\L$~ is not reducible w.r.t. the other roots,
i.e., (\ref{redd}{b,c,d}) are not fulfilled, then ~$F^\L$~ is
irreducible and is isomorphic to the irrep ~$L_\L$~ with this weight.

Other situations shall be discussed below in the context of unitarity.

\subsection{Realization of $\ osp(1\vr\, 2n)\ $ and $\ osp(1\vr\, 2n,\bbr)$}

\nt
The superalgebras ~$osp(m\vr\, 2n) ~=~ osp(m\vr\, 2n)_\o + osp(m\vr\, 2n)_\I$~
are defined as follows \cite{Kab}:
$$osp(m\vr\, 2n)_s ~=~ \{ X\in gl(m/2n;\bbc)_s ~:~ X\ W ~+~ i^{s}\ W\ ^tX = 0
\} \ , \quad s=\o,\I \ ,$$
where ~$W$~ is a matrix of order $m+2n$:
$$ W ~=~ \left( \begin{array}{ccc} iI_m & 0 & 0 \cr
 0 & 0 & I_n \cr
         0 & -I_n & 0 \end{array} \right) $$

The even part ~$osp(m\vr\, 2n)_\o$~ consists of matrices ~$X$~ such that:
\eqnn{evn}
X ~&=&~ \left( \begin{array}{ccc}
S & 0 & 0 \cr
 0 & B & C \cr
 0 & D & -\,{^tB} \end{array} \right)
         \\          \nn\\
&&{^tS}=-S , ~{^tC}=C , ~ {^tD}=D\ .\nn \eea
In our case ~$m=1$~ and ~$S ~=~ 0$.
The Cartan subalgebra ~$\ch$~ consists of diagonal
matrices ~$H$~ such that:
$$ H ~=~ \left( \begin{array}{ccc}
0 & 0 & 0 \cr
 0 & B & 0 \cr
 0 & 0 & -B \end{array} \right)
         $$
We take the following basis for the Cartan subalgebra:
\eqnn{cartb}
H_i ~&=&~ \left( \begin{array}{ccc}
0 & 0 & 0 \cr
 0 & B_i & 0 \cr
 0 & 0 & -B_i \end{array} \right)
\ ,\qquad i<n \ ,               \nn\\
H_n ~&=&~ \left( \begin{array}{ccc}
0 & 0 & 0 \cr
 0 & I_n & 0 \cr
 0 & 0 & -I_n \end{array} \right)
         \eea
where
$$B_i = {\rm diag} (0, \dots 0,1, -1, 0, \dots , 0)$$
the first non-zero entry being on the ~$i$-th place.
This basis shall be used also for the real form
~$osp(1\vr\, 2n,\bbr)$~ and is chosen to be
consistent with the fact that the even subalgebra ~$sp(2n,\bbr)$~
of the latter has as maximal noncompact subalgebra the algebra
~$sl(n,\bbr)\oplus \bbr$. Via the Weyl unitary trick this
related to the structure of ~$sp(2n,\bbr)$~
as a hermitean symmetric space with
maximal compact subalgebra ~$u(n) \cong su(n) \oplus u(1)$.

The root vectors of the roots ~$\d_i - \d_j, ~(i \neq j)$,
~$\d_i + \d_j, ~(i \leq j)$,
~$-(\d_i + \d_j), ~(i \leq j)$,
respectively, are denoted
~$X_{ij}\,$, ~$X^+_{ij}\,$, ~$X^-_{ij}\,$, respectively.
The latter are given by matrices of the type (\ref{evn}) with $S=0$,
given (up to multiplicative normalization) by ~$B = E_{ij},
~C=E_{ij} + E_{ji}$,
~$D = E_{ij} + E_{ji}$, respectively, where ~$E_{ij}$~ is
~$n\times n$~ matrix which has only one non-zero entry equal
to 1 on the intersection of the ~$i$-th row and ~$j$-th column.
Explicitly, (including some choice of normalization), this is:
\eqnn{expl}
X_{ij} ~&=&~ \left( \begin{array}{ccc}
0 & 0 & 0 \cr
 0 & E_{ij} & 0 \cr
 0 & 0 & -E_{ji} \end{array} \right)
         \ , \quad i\neq j\ ,
                 \\ \nn\\
X^+_{ij} ~&=&~ \left( \begin{array}{ccc}
0 & 0 & 0 \cr
 0 & 0 & -E_{ij}- E_{ji} \cr
 0 & 0 & 0 \end{array} \right)
                 \ , \quad i< j \ , \qquad
X^+_{ii} ~=~ \left( \begin{array}{ccc}
0 & 0 & 0 \cr
 0 & 0 & -E_{ii} \cr
 0 & 0 & 0 \end{array} \right)
                \nn\\ \nn\\
X^-_{ij} ~&=&~ \left( \begin{array}{ccc}
0 & 0 & 0 \cr
 0 & 0 & 0 \cr
 0 & E_{ij}+ E_{ji} & 0 \end{array} \right)
                 \ , \quad i< j\ , \qquad
X^-_{ii} ~=~ \left( \begin{array}{ccc}
0 & 0 & 0 \cr
 0 & 0 & 0 \cr
 0 & E_{ii} & 0\end{array} \right)
\nn\eea

The odd part ~$osp(m\vr\, 2n)_\I$~ consists of matrices ~$X$~ such that:
$$ X = \left( \begin{array}{ccc}
0 & \x & -\y \cr
 {^t\y} & 0 & 0 \cr
 {^t\x} & 0 & 0 \end{array} \right)
         $$
The root vectors ~$Y^+_i\,$, ~$Y^-_i\,$,
of the roots ~$\d_i\,,-\d_i$~ correspond to
~$\y$, $\x$, resp., with only non-zero ~$i$-th entry.
Explicitly this is:
\eqnn{expll}
Y^+_{i} ~&=&~ \left( \begin{array}{ccc}
0 & 0 & -E_{1i} \cr
 E_{i1} & 0 & 0 \cr
 0 & 0 & 0 \end{array} \right)
\nn\\ \nn\\
Y^-_{i} ~&=&~ \left( \begin{array}{ccc}
0 & E_{1i}& 0 \cr
 0 & 0 & 0 \cr
 E_{i1} & 0 & 0 \end{array} \right)
    \eea

In the calculations we need all commutators of the kind
~$[X_\b\,,\,X_{-\b}]    ~=~ H_\b\,$, ~$\b\in\D^+_\o\,$. Explicitly,
we have:
\eqna{comh}
[X_{ij}\,,\,X_{ji}] ~&=&~ H_{ij} ~=~ H_i + H_{i+1}
+ \cdots + H_{j-1}~, \quad 1\leq i<j \leq n\ ,
\\
{[Y^+_{i}\,,\,Y^-_{i}]}_+ ~&=&~ H'_{i}
~\equiv~ \left( \begin{array}{ccc}
0 & 0 & 0 \cr
 0 & E_{ii} & 0 \cr
 0 & 0 & -E_{ii} \end{array} \right)
                 ~, \quad 1\leq i \leq n\ ,
\\
{[X^+_{ij}\,,\,X^-_{ij}]} ~&=&~ H'_{ij} ~=~ -H'_i-H'_j
~, \quad 1\leq i<j \leq n \ ,
\\
{[X^+_{ii}\,,\,X^-_{ii}]} ~&=&~ -H'_{i}
~, \quad 1\leq i \leq n\ .
\eena
The minus sign in (\ref{comh}{d}) is consistent with the relations:
\eqn{squar} \half {[Y^\pm_{i},Y^\pm_{i}]}_+ ~=~ (Y^\pm_{i})^2 ~=~
X^\pm_{ii} \ . \end{equation}
We note also the following relations:
\eqnn{comz}
{[Y^+_{i}\,,\,Y^-_{j}]}_+ ~&=&~ X_{ij} \ , \quad i\neq j \ ,
\\
{[Y^\pm_{i}\,,\,Y^\pm_{j}]}_+ ~&=&~ X^\pm_{ij} \ , \quad i\neq j \ ,
\nn\\
H_n ~&=&~ H'_1 + \cdots + H'_n \ .\nn \eea

We shall use also the abstract defining relations of ~$osp(1\vr\, 2n)$ through
the Chevalley basis. Let ~$\hH_i\,$, $i=1,...,n$, be the basis of
the Cartan subalgebra $\ch$ associated with the simple roots, and
~$X^\pm_i\,$, $i=1,...,n$, be
the simple root vectors (the Chevalley generators). The
connection with the basis above is:
\eqnn{relb}
\hH_i ~&=&~ H_i \ , \quad i<n \ , \qquad
\hH_n ~=~ H'_n \ , \nn\\
X^+_i ~&=&~ X^+_{i,i+1} \ , \quad i<n \ , \qquad
X^+_n ~=~ Y^+_n \ . \eea

Let $A = (a_{ij})$ be the Cartan
matrix \cite{Kab}:
\eqn{cbss}
A ~=~ \left( a_{ij} \right) ~~=~~
\left( \begin{array}{ccccccc}
2 & -1 & 0 & \dots &0 & 0 & 0\cr
-1 & 2 & -1 & \dots &0& 0   & 0 \cr
0 & -1 & 2& \dots &0 &  0 & 0 \cr
\dots & \dots & \dots & \dots& \dots & \dots & \dots \cr
0 & 0 & 0 & \dots & 2 & -1 & 0\cr
0 & 0 & 0 & \dots &-1 & 2 & -1 \cr
0 & 0 & 0 & \dots &0    & -2    & 2 \end{array} \right)
\end{equation}
We shall also use the decomposition:
~$A = A^d A^s$, where ~$A^d =$~diag $(1,...,1,2)$,
and ~$A^s$ is a symmetric matrix:
\eqn{cbsss}
A^s ~=~ \left( a^s_{ij} \right) ~~=~~
\left( \begin{array}{ccccccc}
2 & -1 & 0 & \dots &0 & 0 & 0\cr
-1 & 2 & -1 & \dots &0& 0   & 0 \cr
0 & -1 & 2& \dots &0 &  0 & 0 \cr
\dots & \dots & \dots & \dots& \dots & \dots & \dots \cr
0 & 0 & 0 & \dots & 2 & -1 & 0\cr
0 & 0 & 0 & \dots &-1 & 2 & -1 \cr
0 & 0 & 0 & \dots &0    & -1    & 1 \end{array} \right)
\end{equation}

 Then the defining relations of ~$osp(1\vr\, 2n)$~ are:
\eqnn{suco}
&[ \hH_i~, ~ \hH_j] ~=~ 0 ,
~~~[ \hH_i~, ~X^\pm_j] ~=~ \pm a^s_{ij}~ X^\pm_j \ , \nn\\
&[ X^+_i~, ~ X^-_j] ~=~ \d_{ij}~ \hH_i \ ,\\
&({\rm Ad}\, X^\pm_j)^{n_{jk}}(X^\pm_k) ~=~ 0 ~, ~~ j\neq k ~,
~\qquad n_{jk} = 1 - a_{jk} \ , \nn\eea
where in (\ref{suco}) one uses the supercommutator:
\eqnn{supc}
({\rm Ad}\,X^\pm_j)~(X^\pm_k) ~&=&~ [X^\pm_j ,
X^\pm_k] ~\equiv~ \nn\\ &\equiv& ~X^\pm_j X^\pm_k ~-~
(-1)^{ \deg X^\pm_j\ \deg X^\pm_k}
~X^\pm_k X^\pm_j\ . \eea

\subsection{Shapovalov form and unitarity}

\nt
The Shapovalov form is a bilinear $\bbc$--valued form on
~$U(\cg^+)$~ \cite{Sha}, which we extend in the obvious way to
Verma modules, cf. e.g., \cite{DPp}.
We need also the involutive antiautomorphism $\om$ of ~$U(\cg)$~
which will provide the real form we are interested in.
Since this is the split real form ~$osp(1\vr\, 2n,\bbr)$~ we use:
\eqn{omm} \om ( X_\b ) ~=~ X_{-\b} \ , \qquad \om (H) ~=~ H \ ,
\end{equation}
where ~$X_{\b}$~ is the root vector corresponding to the root
~$\b$, ~$H\in\ch$.

Thus, an adaptation of the Shapovalov form suitable for our purposes
is defined as follows:
\eqnn{shh}
&\(\ u\ ,\ u'\ \) ~~=~~
\(\ p\ v_0 \ ,\ p'\ v_0\ \) ~~\equiv~~
\(\ v_0 \ ,\ \om(p)\ p'\ v_0\ \)
~=~ \(\ \om(p')\ p\ v_0 \ , \ v_0\ \) ~,
\\
&u ~=~ p\ v_0\ , ~ ~u' ~=~ p'\ v_0\ , \qquad p,p'\in U(\cg^+),
~~u,u' \in V^\L\ , \nn\eea
supplemented by the normalization condition ~$(v_0, v_0) ~=~ 1$.
 The norms squared of the states would be denoted by:
\eqn{nrm} \, \|\, u \, \|^2 ~\equiv~ \(\, u\, ,\, u\, \) \ .
\end{equation}
Now we need to introduce a PBW basis of ~$U(\cg^+)$.
We use the so-called normal ordering, namely, if we have the relation:
$$ \b ~=~ \b'+ \b'' \ , \quad \b,\b',\b''\in \D^+ $$
then the corresponding root vectors are ordered in the PBW basis
as follows:
\eqn{order} ...\ \(X^+_{\b'}\)^{k'}\
...\ \(X^+_{\b}\)^{k}\
...\ \(X^+_{\b''}\)^{k''}\ ... \ , \qquad k,k',k'' \in \bbz_+ \ .
\end{equation}

We have also to take into account the relation (\ref{squar}) between
the root vectors corresponding to the roots ~$\d_i$~ and ~$2\d_i\,$.
Because of this relation and consistently with (\ref{order}) the
generators ~$X^+_{ii}\,$, $i=1,...,n$, ~are not present in the
PBW basis. On the other hand the PBW basis of the even subalgebra
of ~$U(\cg^+)$~ would differ from the above only in the fact that
the powers of ~$X^+_{i}\,$, $i=1,...,n$, are only even
representing powers of the even generators ~$X^+_{ii}\,$,
$i=1,...,n$.

\section{Unitarity}

\subsection{Calculation of some norms}

\nt
In this subsection we show how to use the form (\ref{shh}) to
calculate the norms of the states.
We shall use the isomorphism between the Cartan subalgebra
~$\ch$~ and its dual ~$\ch^*$. This is given by the
correspondence: to every element ~$\b\in\ch^*$~ there is unique
element ~$H_\b\in\ch$, so that:
\eqn{corr} \mu (H_\b) ~=~ (\mu , \b^\vee ) \ , \end{equation}
for every ~$\mu\in\ch^*$, $\mu\neq 0$. Applying this to the
positive roots
we have: to ~$\b = \d_i-\d_j\,,\d_i\,,\d_i+\d_j\,$, resp., correspond:
~$H_\b = H_{ij}\,,H'_i\,,H'_i+H'_j\,$, resp.

We give now explicitly the norms of the one-particle states
introducing also notation for future use:
\eqnn{norm}
x^-_{ij} ~~&\equiv& ~~ \| \, X_{ij}\, v_0\, \|^2 ~=~
\(\, X_{ij} \ v_0 \ ,\ X_{ij}\ v_0\, \) ~=\nn\\
&=&\ \(\ v_0 \ ,\ X_{ji}\,X_{ij}\ v_0\ \) \ = \
\(\ v_0 \ ,\ (X_{ij}\,X_{ji}-H_{ij}) \ v_0\ \) ~=\nn\\
&=&\ - \L (H_{ij}) ~=~ - (\L, (\d_i-\d_j)^\vee)
~=~ a_i + \cdots + a_{j-1} \ , \quad i<j \ , \\
x^+_{ij} ~~&\equiv& ~~ \| \, X^+_{ij} \, v_0\, \|^2 ~=~
\(\ X^+_{ij} \ v_0 \ ,\ X^+_{ij}\ v_0\ \) ~=\nn\\
&=&\ \(\ v_0 \ ,\ X^-_{ij}\,X^+_{ij}\ v_0\ \) \ = \
\(\ v_0 \ ,\ (X^+_{ij}\,X^-_{ji}-H'_{ij})\ v_0\ \) ~=~ - \L
(H'_{ij}) ~=\nn\\
&=&\ \L (H'_i+H'_j) ~=~ (\L, (\d_i+\d_j)^\vee)
~=~ 2d + a_1 + \cdots + a_{i-1} - a_j - \cdots - a_{n-1}
\ , \nn\\
x_{i} ~~&\equiv& ~~ \| \, X^+_{i} \, v_0\, \|^2 ~=~
\(\ X^+_{i} \ v_0 \ ,\ X^+_{i}\ v_0\ \) ~=\nn\\
&=&\ \(\ v_0 \ ,\ X^-_{i}\,X^+_{i}\ v_0\ \) \ = \
\(\ v_0 \ ,\ (- X^+_{i}\,X^-_{i} + H'_{i})\ v_0\ \) ~= \L (H'_{i}) ~=\nn\\
&=&\ (\L, \d_i^\vee)
~=~ 2d \ +\ a_1 + \cdots + a_{i-1} - a_i - \cdots - a_{n-1} \ .\nn
\eea

Positivity of all these norms gives the following necessary
conditions for unitarity:
\eqnn{posi}
 a_i ~&\geq& ~0 \ , \qquad i=1,...,n-1\ , \nn\\
 d ~&\geq& ~\half (a_1 + \cdots + a_{n-1})\ . \eea
In fact, the boundary values are possible due to factoring out of
the corresponding null states when passing from the Verma module
to the unitary irreducible factor module.

Further, we shall discuss only norms which involve the conformal
weight since the others are related to unitarity of the irrep
restricted to the maximal simple compact subalgebra $su(n)$.
The norms that we are going to consider can be written in terms
of factors ~$(d ~-~ ...)$, and the leading term in ~$d$~ has a
positive coefficient. Thus, for ~$d$~ large enough all norms will
be positive. When ~$d$~ is decreasing there is a critical point
at which one (or more) norm(s) will become zero. This critical
point (called the 'first reduction point' in \cite{FF}) can be read off
from the reducibility conditions, since at that point the Verma
module is reducible (and it is the corresponding submodule that
has zero norm states).

The maximal ~$d$~ coming from the different possibilities in
(\ref{redd}{b}) are obtained for $m^+_{ij}=1$ and they are, denoting
also the corresponding root:
\eqn{boun}
d_{ij} ~\equiv~ n + \half(a_j + \cdots + a_{n-1}
- a_1 - \cdots - a_{i-1}-i-j) \ ,
\end{equation}
the corresponding root being ~$\d_i+\d_j\,$.
The maximal ~$d$~ coming from the different possibilities in
(\ref{redd}{c,d}), resp., are obtained for $m_{i}=1$, $m_{ii}=1$,
resp., and they are:
\eqnn{bounz}
&&d_{i} ~\equiv~ n -i + \half(a_i + \cdots + a_{n-1}
- a_1 - \cdots - a_{i-1}) \ ,
\\
&&d_{ii} ~=~ d_i - \half \ ,
\nn\eea
the corresponding roots being ~$\d_i\,,2\d_j\,$, resp.
These are some orderings between these maximal reduction points:
\eqnn{comp}
d_1 ~&>&~ d_2 ~>~ \cdots ~>~ d_n \ , \\
d_{i,i+1} ~&>&~ d_{i,i+2} ~>~ \cdots ~>~ d_{in} \ , \nn\\
d_{1,j} ~&>&~ d_{2,j} ~>~ \cdots ~>~ d_{j-1,j} \ , \nn\\
d_i ~&>&~ d_{jk} ~>~ d_\ell
\ , \qquad i\leq j <k \leq \ell \ . \nn\eea

Obviously the first reduction point is:
\eqn{frp} d_{1} ~=~ n -1 + \half(a_1 + \cdots + a_{n-1})  \ .
\end{equation}

\subsection{Main result}

\nt {\bf Theorem:}~~ All positive energy unitary irreducible
representations of the superalgebras ~$osp(1\vr\, 2n,\bbr)$~ characterized
by the signature ~$\chi$~ in (\ref{sgn}) are obtained for real ~$d$~
and are given in the following list:
\eqnn{unt}
&&d ~\geq~ d_{1} ~=~ n -1 + \half(a_1 + \cdots + a_{n-1})
\ , \quad   a_1\neq 0 \ , \\
&&d ~\geq~ d_{12} ~=~ n - 2 + \half(a_2 + \cdots + a_{n-1} +1)
\ , \quad a_1 = 0,\  a_2\neq 0\ , \nn\\
&& ... \nn\\
&&d ~\geq~ d_{j-1,j} ~=~ n -j + \half(a_j + \cdots + a_{n-1} +1)
\ , \quad a_1 = ... = a_{j-1} = 0 ,\  a_j\neq 0\ , \nn\\
&& ... \nn\\
&&d ~\geq~ d_{n-1,n} ~=~ \half \ , \quad a_1 ~=~ ... ~=~ a_{n-1} ~=~
0 \ .\nn \eea {\bf Sketch of Proof:}~~ \nl   The statement of the
Theorem for ~$d ~>~ d_1$~ is clear form the general considerations
above. For ~$d ~=~ d_1$~ we have the first zero norm state which is
naturally given by the corresponding singular vector ~$v^{1,\d_1}
~=~ \cp^{1,\d_1}\ v_0\,$. In fact, all states of the embedded
submodule ~$V^{\L+\d_1}$~ built on ~$v^{1,\d_1}$~ have zero norms.
Due to the above singular vector we have the following additional
null condition in ~$F^\L$~: \eqn{nula} \cp^{1,\d_1}\ \witl ~=~ 0 \ .
\end{equation} The above conditions factorizes the submodule built
on ~$v^{1,\d_1}\,$. There are no other vectors with zero norm at
$d=d_1$ since by a general result \cite{Kc}, the elementary
embeddings between Verma modules are one-dimensional. Thus, ~$F^\L$~
is the UIR ~$L_\L ~=~ F^\L$.

Further we consider the remaining discrete points of unitarity
for ~$d ~<~ d_1\,$, i.e., ~$d ~=~ d_{i,i+1}\,$, $i=1,...,n-1$.
The corresponding roots are ~$\d_i+\d_{i+1} ~=~
\a_i + 2\a_{i+1} +\cdots + 2\a_n\,$.
The corresponding singular vectors ~$v^{1,\d_i+\d_{i+1}}
~=~ \cp^{1,\d_i+\d_{i+1}}\ v_0\,$.

Now, fix ~$i$, where $i\in \{1,...,n-1\}$.
All states of the embedded submodule
~$V^{\L+\d_i+\d_{i+1}}$~ built on ~$v^{1,\d_i+\d_{i+1}}$~ have
zero norms for ~$d ~=~ d_{i,i+1}\,$.
Due to the above singular vector we have the following
additional null condition in ~$F^\L$~:
\eqn{nuli} \cp^{1,\d_i+\d_{i+1}}\ \witl ~=~ 0 \ ,
\qquad d ~=~d_{i,i+1}\, . \end{equation}
At this point the states built on the
vector ~$v^{1,\d_1}$~ and on the vectors
~$v^{1,\d_k+\d_{k+1}}$~ for $k<i$
(all of these are not singular vectors at
$d ~=~ d_{i,i+1}$) have negative norm except
when ~$a_1=\cdots = a_i =0$. For this statement we may use
the explicit form of these vectors. This explicit form is
the same as the singular vectors of the same weight for the Lie
algebra ~$B_n ~=~ so(2n+1)$. For ~$v^{1,\d_1}$~ this
can be read off from \cite{Dos} (in fact, there it is for the more
general situation of the quantum group ~$U_q(B_n)$)~:
\eqna{singg}
v^{1,\d_1}
~&=&~ \sum_{k_1=0}^1 \cdots \sum_{k_{n-1}=0}^1\ b_{k_1\dots
k_{n-1}}\ (X^+_1)^{1-k_1} \cdots (X^+_{n-1})^{1-k_{n-1}}\
\times \nn\\ &\times&\
X^+_n\ (X^+_{n-1})^{k_{n-1}} \cdots\
(X^+_1)^{k_1}\ v_0\ \equiv \cp^{1,\d_1}\ v_0 \ ,
\\ \nn\\
 b_{k_1\dots k_{n-1}} ~&=&~
(-1)^{k_1 + \cdots + k_{n-1}} \ b_0
\ \frac{(\r - \L)(H^1)}{(\r - \L)(H^1) - k_1}
\cdots \frac{(\r - \L)(H^{n-1})}{(\r - \L)(H^{n-1}) - k_{n-1} }
\ = \\
&=&\ (-1)^{k_1 + \cdots + k_{n-1}}
\ b_0
\ \frac{ 1+a_1}{ 1+a_1 - k_1} \cdots
\frac{n-1 +a_1+\cdots +a_{n-1} }{
n-1+a_1 +\cdots +a_{n-1} - k_{n-1}} \ = \\
&=&\ (-1)^{k_1 + \cdots + k_{n-1}}\ (a_1 +k)\
\frac{ 2+a_1+a_2 }{ 2+a_1+a_2 - k_2} \cdots
\frac{n-1 +a_1+\cdots +a_{n-1} }{
n-1+a_1 +\cdots +a_{n-1} - k_{n-1}} \qquad\qquad
\eena
where \ $H^s
\ =\ \hH_1 + \hH_2 + \cdots + \hH_s$\ , (cf. f-lae (13) from \cite{Dos} with
$q=1$, $t=n-1$, $m=1$, $\l\to -\L$ (the last change due to the
fact that in \cite{Dos} are considered highest weight modules));
in (\ref{singg}{c}) we have inserted our signatures:
$$ (\r - \L)(H^s) ~=~ (\r - \L, (\a_1+\cdots +\a_s)^\vee)
~=~ (\r - \L, \d_1-\d_{s+1} ) = m^-_{1,s+1} ~=~ s + a_1 + \cdots +
a_s $$ and in (\ref{singg}{d}) we have made the choice of constant
~$b_0 = a_1$~ in order to make the expression valid also for
~$a_1=0$. It is easy to see that for ~$a_1=0$~ the vector
~$v^{1,\d_1}$~ is not independent, but is a descendant of the
singular vector ~$v^1_s ~=~ X^+_1\, v_0$~: \eqnn{spes} v^{1,\d_1}
~&=&~ \sum_{k_2=0}^1 \cdots \sum_{k_{n-1}=0}^1\ b_{1,k_2\dots
k_{n-1}}\ (X^+_2)^{1-k_2} \cdots (X^+_{n-1})^{1-k_{n-1}}\ \times
\nn\\ &&\times\ X^+_n\ (X^+_{n-1})^{k_{n-1}} \cdots\ (X^+_2)^{k_2}\
X^+_1 \ v_0 \ .\eea Thus, ~$v^{1,\d_1}$~ is not present in ~$F^\L$~
for any ~$d$~ and ~$a_1 ~=~ 0$~ since the null condition
(\ref{nula}) follows from case ~$i=1$~ of the null conditions
(\ref{nullm}). Analogously, if $i>1$ and fixing now $k<i$, the
vector ~$v^{1,\d_k+\d_{k+1}}$~ has negative norm at ~$d
~=~d_{i,i+1}$~ except if ~$a_{k+1} ~=~ 0$, when it is not
independent, but is a descendant of the singular vector ~$v^{k+1}_s
~=~ X^+_{k+1}\ v_0\,$, and hence is not present in ~$F^\L$.   Thus,
for ~$d ~=~ d_{i,i+1}$~ together with ~$a_1 ~=~ \cdots ~=~ a_{i} ~=~
0$, the condition (\ref{nuli}) factorizing the submodule built on
~$v^{1,\d_i+\d_{i+1}}\,$, is  needed together wuth (\ref{nullm})  to
obtain the UIR ~$L_\L ~=~ F^\L$~ at
 ~$d ~=~ d_{i,i+1}\,$, ~$i=1,...,n-1$. The complete Proof should appear in
\cite{DoSa}. \hfill \bsq

\bigskip

\section*{Acknowledgements} VKD would like to thank S.
Ferrara for attracting his attention to the problem and for
discussions and the Math. Dept. of Sydney Univ.  (where this work was
started) for hospitality in September-October 2002. Both authors would like to
thank Jouko Mickelsson and Harald Grosse for inviting them to participate in
the programme "Noncommutative Geometry and Quantum Field Theory",
October-November 2002, ESI, Vienna (where part of the work was done).
V.K.D. was supported in part by the Bulgarian National Council for
Scientific Research grant F-1205/02.

\np

\end{document}